\documentclass[aps,twocolumn,showpacs,superscriptaddress]{revtex4}
\usepackage {graphicx}

\begin{document}
\title{Magnetoresistive study of antiferromagnetic--weak ferromagnetic
transition in single-crystal La$_{2}$CuO$_{4+\delta}$ }

\author{B. I. Belevtsev}
\email[]{belevtsev@ilt.kharkov.ua}
%\homepage[]{Your web page}
%\thanks{}
%\altaffiliation{}
\affiliation{B. Verkin Institute for Low Temperature Physics and
Engineering, National Academy of Sciences, pr. Lenina 47, Kharkov
61103, Ukraine}

\author{N. V. Dalakova}
\affiliation{B. Verkin Institute for Low Temperature Physics and
Engineering, National Academy of Sciences, pr. Lenina 47, Kharkov
61103, Ukraine}

\author{V. N. Savitsky}
\affiliation{B. Verkin Institute for Low Temperature Physics and
Engineering, National Academy of Sciences, pr. Lenina 47, Kharkov
61103, Ukraine}

\author{A. V. Bondarenko}
\affiliation{V. N. Karazin Kharkov National University, pl.
Svobodi 4, Kharkov, 61077, Ukraine}

\author{A. S. Panfilov}
\affiliation{B. Verkin Institute for Low Temperature Physics and
Engineering, National Academy of Sciences, pr. Lenina 47, Kharkov
61103, Ukraine}

\author{I. S. Braude}
\affiliation{B. Verkin Institute for Low Temperature Physics and
Engineering, National Academy of Sciences, pr. Lenina 47, Kharkov
61103, Ukraine}

\begin{abstract}
The resistive measurements were made to study the magnetic
field-induced antiferromagnetic (AF) - weak ferromagnetic (WF)
transition in  La$_2$CuO$_4$  single-crystal. The magnetic field
(DC or pulsed) was applied normally to the  CuO$_2$  layers. The
transition manifested itself in a drastic decrease of the
resistance in critical fields of ~5-7 T. The study is the first to
display the effect of the AF -WF transition on the conductivity of
the La$_2$CuO$_4$ single-crystal  in the parallel - to - CuO$_2$
layers direction. The results provide support for the
3-dimensional nature of the hopping conduction of this layered
oxide.
\end{abstract}

\pacs{74.72.Dn; 75.30.Kz; 75.50.Ee}
\maketitle

\section{Introduction}

Transport and magnetic properties of cuprate
La$_{2}$CuO$_{4+\delta}$ attract up to now a considerable
attention in the area of superconducting research. This is a
parent compound for one of the family of high-temperature
superconductors, and study of its properties is considered to be
important for elucidation of still unclear nature of
superconductivity in cuprates. Stoichiometric La$_{2}$CuO$_{4} $
($\delta = 0$) is an antiferromagnetic (AF) insulator with
N\'{e}el temperature, $T_{N} $, about 320 K, but doping it with
bivalent metals (such as Sr) or with excess oxygen ($\delta \ne
0$) leads to violation of long-range AF order and decrease in
$T_{N}$ \cite{aharon,endoh,kastner}. A fairly high doping results
in transition to metallic state.
\par
Perovskite crystal lattice of La$_{2}$CuO$_{4}$ is an orthorhombic
(below about 530 K) consisting of CuO$_{2}$-layers separated by
La$_{2}$O$_{2}$ layers (the latter consisting of two buckled La-O
layers) \cite{endoh,kastner}. In \textit{Bmab} space group, the
CuO$_{2}$ layers are perpendicular to the \textbf{\textit{c}} axis
and parallel to \textbf{\textit{a-b}} plane \cite{kastner}. The
CuO$_{6}$ octahedra are tilted in a staggered way; the tilting is
uniform in a given \textbf{\textit{c-b}} plane. The AF state is
strongly connected with crystal lattice features \cite{vaknin}.
The magnetic state is determined by $d^{9}$Cu$^{2+}$ ions with
spin $S =0.5$. In the CuO$_{2}$ planes, the magnetic structure is
characterized by simple two-dimensional (2D) AF array with nearest
neighbors having antiparallel moments \cite{vaknin}. Due to the
above-mentioned tilting of the CuO$_{6}$ octahedra, the spins are
canted 0.17$^{\circ}$ in the \textbf{\textit{c-b}} plane away from
the \textbf{\textit{b}} axis \cite{vaknin,thio}. As a result, a
weak ferromagnetic (FM) moment perpendicular to CuO$_{2}$ plane
appears in each layer. Below $T_{N}$, the directions of FM
moments are opposite in neighboring  CuO$_{2}$ planes, so that the
system as a whole is three-dimensional (3D) AF \cite{thio}.
\par
Application of high enough magnetic field along the
\textbf{c-}axis causes a magnetic transition into a
weak-ferromagnetic (WF) state, in which all canted moments are
aligned along the field direction \cite{thio}. The transition is
accompanied by a jump-like change in the resistivity \cite{thio}.
The critical field $H_{c}$ of the transition is temperature
dependent. It goes to zero for $T$ approaching $T_{N}$, but
increases with temperature decreasing and amounts up to 5-6 T
below 100 K. Hole doping of La$_{2}$CuO$_{4}$ leading to smaller
$T_{N}$ causes smaller $H_{c}$ values as well (down to about 3 T
at low temperature for samples with $T_{N}$ about 100 K)
\cite{bazhan}. Some magnetic transitions have also been found for
field applied parallel to \textbf{\textit{a-b}} plane
\cite{thio2}. In this case, for field parallel to
\textbf{\textit{b}} axis, a spin-flop transition was found at
field $H_{1}$ about 10 T, and a transition to FM state at field
$H_{2}$ about 20 T. These transitions manifest themselves as weak
knees (no jumps) in MR curves \cite{thio2}.  It is believed that
no magnetic transition should take place when field is applied
parallel to axis \textbf{\textit{a}}, which is perpendicular to
the staggered moments \cite{thio2,gogolin}.
\par
The doping with excess oxygen introduces charge carriers (holes)
in CuO$_{2}$ planes. At small enough $\delta$ ($< 0.01$),
La$_{2}$CuO$_{4+\delta}$ remains insulating, even if $T_{N}$
reducing considerably \cite{zakhar,statt}. The excess oxygen atoms
reside at an interstitial sites between LaO planes \cite{chaill}.
Each of such excess atom is surrounded by tetrahedron of apical
oxygen atoms. For layered cuprates, in which the CuO$_{2}$ planes
are the main conducting units, a quasi 2D behavior to be expected
for the in-plane transport. This has really been found in many
cuprates \cite{iye}, but not in La$_{2}$CuO$_{4+\delta}$. In this
compound, the Mott's variable-range hopping (VRH) with
temperature dependence of resistance described by expression,

\begin{equation}
\label{eq1} R \approx R_{0} \exp\left( {\frac{{T_{0}} }{{T}}}
\right)^{1/4},
\end{equation}

\noindent is found \cite{kastner2,boris1} at low $T$ for both, the
in-plane (current $J$ parallel to CuO$_{2}$ planes) and
out-of-plane ($J\|$\textbf{\textit{c}}) transport. Fractional
exponent in Eq. (\ref{eq1}) equal to 1/4 corresponds to 3D system
(for 2D systems, it should be equal to 1/3) \cite{shklov}. At the
same time, the hopping conduction in La$_{2}$CuO$_{4+\delta}$
samples with fairly high crystal perfection shows a considerable
anisotropy, so that magnitudes of $R_{0} $ and $T_{0}$ in Eq.
(\ref{eq1}) are different for the in-plane and out-of-plane
transport. The in-plane conductivity $\sigma_{ab}$ is found to be
considerably higher than the out-of-plane one $\sigma_{c}$. A
ratio $\sigma_{ab} /\sigma_{c}$ is strongly temperature dependent.
It is minimal (about 10) in liquid-helium temperature range, but
increases dramatically with temperature and saturates above 200 K
to a maximal values order of 100 \cite{preyer,hundley,boris2}.
\par
The 3D character of VRH in La$_{2}$CuO$_{4+\delta}$ testifies that
a hole transfer between CuO$_{2}$ is likely not only at
$J\|$\textbf{\textit{c}}, but at $J\|$\textbf{\textit{a,b}} as
well. At considering this question it is important to know the
exact nature of holes in La$_{2}$CuO$_{4+\delta}$. Although about
17 years has passed since the discovery of superconductivity in
doped La$_{2}$CuO$_{4+\delta}$, the nature of holes in it still
cannot be considered as completely clear. This in turn hampers to
gain an insight into the nature of cuprate's superconductivity. In
the undoped state, CuO$_{2}$ planes present a lattice of
$d^{9}$Cu$^{+2}$ ($S = 0.5$) and $p^{6}$O$^{-2}$ ($S = 0$) ions.
Doping with excess oxygen causes (to ensure neutrality)
appearance of additional holes in the planes. This can be
achieved in two ways: 1) some of the $d^{9}$Cu$^{+2}$ ions change
into $d^{8}$Cu$^{+3}$ ($S = 0$) state, or 2) some of the in-plane
oxygen ions $p^{6}$O$^{-2}$ change into $p^{5}$O$^{-1}$ ($S =
0.5$) state. In either case, the holes induce strong local
perturbations of AF order. In known literature
\cite{emery,pickett,kremer,brenig,loktev,yeh,carlson}, both kinds
of holes have been taken into account in theoretical models of
fundamental properties of the cuprates. There is much
speculation, however, that holes in La$_{2}$CuO$_{4+\delta}$ have
a strong oxygen character
\cite{emery,pickett,kremer,brenig,loktev,yeh}, and this view has a
strong experimental support \cite{pickett,brenig,tranquada}. At
the same time, due to overlapping of {\textit{d}}- and
{\textit{p}}-orbitals and hybridization of {\textit{d}}- and
{\textit{p}}-bands, {\textit{d}}-orbitals exert a significant
influence on the hole motion.
\par
According to Ref. \cite{kremer}, owing to special character of the
excess oxygen as interstitial atom \cite{chaill} with weak
oxygen-oxygen bonding, the holes can be delocalized from CuO$_{2}$
planes onto the apical O atoms, i.e. into La$_{2}$O$_{2+\delta}$
region between adjacent CuO$_{2}$ planes. This assures the 3D
nature of VRH in La$_{2}$CuO$_{4+\delta}$. In this way the
La$_{2}$CuO$_{4+\delta}$ differs drastically from Sr doped system,
where the holes remain quasi two-dimensional. Really, the ratio
$\sigma_{ab}/\sigma_{c}$ in lightly doped
La$_{1-x}$Sr$_{x}$CuO$_{4}$ crystals of good quality can be as
high as several thousands \cite{komiya}.
\par
In this communication, we report the results of study of AF-WF
transition by magnetoresistance (MR) measurements in a
La$_{2}$CuO$_{4+\delta}$ single crystal. In known previous studies
\cite{thio,cheong,balaev,zakhar2,zakhar3} the MR investigations of
AF-WF transition in La$_{2}$CuO$_{4+\delta}$ were done for the
case when both the magnetic field and transport current are
perpendicular to the CuO$_{2}$ planes (i.e.,
$H\|$\textbf{\textit{c} }and $J\|$\textbf{\textit{c}}). Under
these conditions a rather sharp decrease in the resistance has
been found as the critical field $H_{c}$ was approached from
below. Amplitude of the relative change in resistance ($\Delta
R/R_{n} $, where $R_{n} $ is the resistance in the AF state) due
to AF-WF transition depends on temperature. It is maximal in the
range 20-30 K, where it can amount up to 0.30-0.50 in fairly
perfect crystals \cite{thio,thio2,balaev,zakhar2,zakhar3}.
\par
It is known that the enhancement of spin order usually leads to
decrease in resistivity of metallic systems. For example, a
considerable decrease in resistivity can occur at transitions from
paramagnetic to FM or AF state in some metals, alloys or even in
some FM perovskite oxides, like mixed-valence manganites
\cite{vonsov,gratz,coey}. This is usually attributed to a decrease
in the scattering rate of quasi-free charge carriers on disordered
local spins as a result of the above-mentioned magnetic
transitions. The situation is rather different in the case of
insulating La$_{2}$CuO$_{4+\delta}$. Here the transition to 3D AF
state produces hardly any noticeable change in the hopping
conductivity at $T_{N}$ (apparently for the reason that 2D AF
correlations in CuO$_{2}$ planes persist up to temperatures far
above $T_{N}$ \cite{aharon,endoh,kastner}). But transition to the
3D WF state increases the conductivity enormously. Since VRH in
La$_{2}$CuO$_{4+\delta}$ has a pronounced 3D character, it can be
expected that AF-WF transition would manifest itself in
resistivity in field $H\|$\textbf{\textit{c}} not only for the
transport current perpendicular to the CuO$_{2}$ planes, as it was
found in Refs. \cite{thio,balaev,zakhar2,zakhar3}, but for the
in-plane hole transport as well. In this study this effect has
been actually revealed, as described below.
\bigskip
\section{Experimental}
\bigskip
A single-crystal La$_{2}$CuO$_{4+\delta}$ sample with dimensions
1.3$\times$0.3$\times$0.39 mm is investigated. This sample was
studied previously in work \cite{boris2}, where it was indicated
as a sample No.~1 with $T_{N}$ = 188 K. After that study, the
sample was annealed additionally in oxygen atmosphere
(700$^{\circ}$C, 5.5 days) in the hope that oxygen content (that
is, $\delta $) will be increased. It turns out, however, that the
thermal treatment has caused only a slight decrease in $T_{N} $
(down to 182 K) and in the resistivity. The $T_{N} $ value was
determined from magnetic susceptibility measurements.
\par
The crystallographic orientation of the sample was determined from
an x-ray diffraction study. This reveals that the sample has a
quantity of twins, which inevitably appear in
La$_{2}$CuO$_{4+\delta}$ crystals when cooled through tetrahedral
to orthorhombic structure transition at $T\approx$~530 K
\cite{kastner}. As a result, a peculiar domain structure is
developed. Orientation of \textbf{\textit{c}}-axis is the same in
each domain, but orientations of \textbf{\textit{a}} and
\textbf{\textit{b}} axes are switched (or reversed) in a fixed way
between two possible orientations upon crossing the domain (twin)
boundaries. In this connection, although we will speak
conventionally in the following about \textbf{\textit{a}}- or
\textbf{\textit{b}}- directions of transport current in the sample
studied, they should be taken, first of all, as the two in-plane
current directions in the twinned crystal, which are perpendicular
to each other. In a heavily twinned crystal no significant
anisotropy in the in-plane conductivity can be expected even
assuming that some intrinsic conductivity anisotropy within
CuO$_{2}$ planes takes place. We have found, however, a
pronounced anisotropy in the conductivity (and a rather
significant one in the MR) for these two in-plane directions.
This matter will be touched upon in the next section of the
paper. In contrast, we can speak about
\textbf{\textit{c}}-directions in the sample studied without any
reservation or possible misunderstanding.
\par
In this study, the DC resistance in the directions parallel to
CuO$_{2}$ planes was measured by the Montgomery method
\cite{mont}, which is appropriate for systems with a pronounced
anisotropy of the conductivity. Contacts between the measuring
wires and the sample were made using a conducting silver paste.
The measurements were done in field $H\|$\textbf{\textit{c}} in a
helium cryostat with a superconducting solenoid. Although maximum
field in the cryostat (about 6 T) has appeared to be quite
sufficient in most cases to reveal manifestation of the AF-WF
transition in MR of the sample studied, a somewhat higher field is
needed to study the transition more thoroughly, especially for
study of hysteretic phenomena in $R\left({H} \right)$ curves in
vicinity of critical field $H_{c}$
\cite{thio,balaev,zakhar2,zakhar3}. This hysteretic behavior is
considered as an indication of the first-order transition. For
this reason, a part of DC resistance measurements in this study
were done in pulse magnetic field with amplitude up to 15 T. A
nearly sinusoidal pulse has a duration about 33 ms, during which
the field is swept from zero to a maximum amplitude and back to
zero. For these measurements the field $H \|$\textbf{\textit{c}}
and transport currents, $J\|$\textbf{\textit{c}} and
$J\|$\textbf{\textit{a}}, were used. The rate of variations in
magnetic field was up to 10$^{3}$ T/s.  Other essential details of
the pulse measuring technique employed can be found in Ref.
\cite{dudko}.

\section{Results and discussion}
The temperature dependences of resistivity $\rho_{a}$, measured
along axis \textbf{\textit{a}} ($J\|$\textbf{\textit{a}}), is
shown in Fig. 1 for different magnitudes of measuring current. It
is seen that $\rho(T)$ behavior does not depend essentially on
current in the whole measuring temperature range, 4.2~K $\leq T
\leq$~300~K, for current magnitude lesser than about 1~$\mu$A;
that is Ohm's law holds in this case. For better consideration,
one of these ohmic $R(T)$ curves (at $J = 1$~$\mu $A) is presented
separately in Fig. 2. It can be seen that Mott's law [Eq.
(\ref{eq1})] is obeyed fairly well in the range 20 K $\lesssim T
\lesssim $ 200~K. In the range $T<20$~K, a steeper [as compared to
Eq. (\ref{eq1})] increase in $R$ with decreasing temperature is
found. This deviation from Mott's law at low temperature is rather
typical for La$_{2}$CuO$_{4+\delta}$, and was observed earlier in
Refs. \cite{boris1,zakhar3}. In Ref. \cite{boris1}, a possible
reason for this behavior is suggested to be determined by the
presence of superconducting inclusions in insulating sample due to
phase separation in La$_{2}$CuO$_{4+\delta}$.
\par
Magnetic structure of La$_{2}$CuO$_{4+\delta}$, according to
neutron diffraction data \cite{endoh,vaknin}, is anisotropic for
all three orthorhombic axes. The same can be expected, therefore,
for transport and magnetic properties. In the presence of twins,
however, the measured transport and magnetic properties show
usually quite definite anisotropy solely for directions parallel
and perpendicular to the CuO$_{2}$ planes. Recently, in untwinned
La$_{2}$CuO$_{4+\delta}$ crystals, a clear in-plane anisotropy of
the magnetic susceptibility $\chi$ was found \cite{lavrov}. A
similar phenomenon may be expected in transport properties of
La$_{2}$CuO$_{4+\delta}$ samples without twins.
\par
In a sample with multiple twins, no considerable in-plane
anisotropy could be expected. The measured ratio $\rho _{b} /\rho
_{a} $ in the sample studied (see inset in Fig. 2) reveals,
however, a rather distinct anisotropy. The ratio is close to unity
at $T \approx $ 11-12 K, but it increases with temperature and
approaches value of about 3 at room temperature. A similar
behavior was found in the previously studied sample with somewhat
higher $T_{N} \approx $ 188 K \cite{boris2}. The
\textbf{\textit{a-b}} anisotropic conductivity behavior in a
twinned sample (in the case that the conductivities $\sigma _{a} $
and $\sigma _{b} $ are inherently different) can be observed only
when, \textit{first}, available twins are few in number (so that
measured resistivity is not properly averaged between the two
possible crystal orientations), and, \textit{second}, a given
current direction is really parallel to axis \textbf{\textit{a}}
(or \textbf{\textit{b}}) in the most part of the crystal. The
results of this study give, therefore, an evidence that the
intrinsic conductivity anisotropy in CuO$_2$ planes of
La$_{2}$CuO$_{4+\delta}$ is quite credible.
\par
We found that MR behavior of the sample studied depends
significantly on magnitude of measuring current, especially at low
temperature. Upper panel of Fig. 3 presents the MR curves recorded
at $T=5$~K for the case $J\|$\textbf{\textit{a}}. It can be seen
that for low currents (that is in ohmic regime) the MR is
positive, but for high enough currents ($J\geq 1 \mu$A) the MR
becomes negative and strongly increases above $H\simeq 5$~T.
Positive MR was observed only at low temperature ($T<20$~K) for
both the in-plane current directions used,
$J\|$\textbf{\textit{a}} and $J\|$\textbf{\textit{b}}. At fairly
high temperature, $T\geq 20$~K, only negative MR is observed,
which increases profoundly above $H\simeq 5$~T, as well (lower
panel of Fig. 3). We have attributed this rather sharp increase
to an influence of the AF-WF transition, as it will be discussed
in more detail below. As for the positive MR at low temperature
($T<20$~K), this could be attributed to the presence of
superconducting inclusions due to phase separation as it was
mentioned above. For example, in Ref. \cite{boris3}, positive MR
attributed to superconducting inclusions has been found at low
temperature range ($T<10$~K) in even more resistive
La$_{2}$CuO$_{4+\delta}$ with higher $T_N$.
\par
For all temperature range, where MR is measured in this study (4.2
K $\leq T \leq 90$~K), the MR magnitude was strongly dependent on
measuring current (as illustrated by Fig.~3). For this reason, to
compare MR curves with an evident effect of AF-WF transition at
different temperatures we have used only data for rather high
currents, that is for non-ohmic conduction regime. Some examples
of the MR curves at $H\|$\textbf{\textit{c}} for the cases
$J\|$\textbf{\textit{a}} and $J\|$\textbf{\textit{b}} and current
$J=100$~$\mu$A are shown in Figs. 4 and 5 for certain selected
temperatures. It is obvious from the curves that a rather sharp
decrease in resistance occurs when $H$ exceeds some critical
magnitude (in the range 5-6 T). All main features of this
resistive transition are quite identical to those found in MR
behavior of La$_{2}$CuO$_{4+\delta}$ at the AF-WF transition for
the case $H\|$\textbf{\textit{c}}, and the out-of-plane current
direction ($J\|$\textbf{\textit{c}}), when mainly inter-plane
hopping is affected by the transition
\cite{thio,cheong,balaev,zakhar2}. The results obtained show that
the AF-WF transition influences hopping conduction in the
directions parallel to CuO$_{2}$ planes as well. This effect,
although being anticipated (as it is indicated above), have never
been seen previously in La$_{2}$CuO$_{4+\delta}$, to our
knowledge.
\par
The following features of the resistive transition can be pointed
out. First, the transition is sharper and the relative changes in
resistance, $\Delta R/R_{n}$, are larger for
\textbf{\textit{b}}-direction of the transport current than those
for \textbf{\textit{a}}-direction (compare Figs. 4 and 5). Second,
the MR curves are hysteretic in the field range of the transition,
as expected. The hysteresis becomes more pronounced for decreasing
temperature. The latter feature of the MR curves is quite
consistent with that found previously at the AF-WF transition for
$J\|$\textbf{\textit{c}} \cite{thio,balaev,zakhar2,zakhar3}.
Third,  a considerable negative MR in low field range below the
magnetic transition can be observed (Figs. 4 and 5). This
contribution to the total MR is not hysteretic and, maybe, has a
little if any relationship to the magnetic transition. For a
given current (for example, for $J=100$~$\mu$A, as in Figs. 4 and
5) the contribution of this type of MR increases with temperature
decreasing and is more pronounced for
\textbf{\textit{a}}-direction of measuring current. It is found
as well that the negative MR at low field increases with current
magnitude (Fig. 3) and, therefore, with an applied voltage, so it
is much more pronounced in non-ohmic regime of hopping
conductivity (compare Fig. 1 and Fig. 3). In previous studies,
negative MR in the AF La$_{2}$CuO$_{4+\delta}$ for the case of
both the current and field parallel to CuO$_{2}$ was found and
discussed to a certain degree \cite{boris1,boris2}. The nature of
the negative MR in rather low field $H\|$\textbf{\textit{c}} and
$J\|$\textbf{\textit{a,b}} revealed in this work is not clear and
is worthy of additional study.
\par
It is evident from Figs. 4 and 5 that the maximal DC field about 6
T, used for these measurements, is not high enough to accomplish
the magnetic transition in full measure. To overcome this
disadvantage, the measurements in pulse magnetic field with
magnitude up to 15 T were done. The MR curves were recorded at
temperatures $T = $4.2 K, 20.4 K and 77~K for both the in-plane
and out-of-plane directions of the transport current. Examples of
MR curves for pulse field at $T = $ 4.2 K and $T=$ 77 K are shown
in Figs. 6 and 7.
\par
The pulse MR measurements have enabled to see the magnetic
transitions in full measure. The MR curves for low temperature
region were found to be quite similar for both methods (compare
Fig. 4 and Fig. 6). It is also seen that resistive transition for
out-of-plane current direction is sharper, and the relative
changes in resistance, $\Delta R/R_{n}, $ are generally larger
than those for in-plane direction.  MR curves in pulse magnetic
field at $T=77$~K are less hysteretic than those at $T=4.2$~K, as
expected (Fig. 7). Maximum values of $\Delta R/R_{n} \approx
50$~\%, found in this study for pulsed magnetic field agree well
with those found in previous studies in DC magnetic field
\cite{thio2}.
\par
In conclusion, we have found that the AF-WF transition in
La$_{2}$CuO$_{4+\delta}$ clearly manifests itself in the in-plane
hopping conductivity. This supports the 3D nature of hopping
conduction in this compound.

\newpage
\centerline{\bf Figure captions}\vspace{12pt}

Figure 1. The temperature dependences of resistivity $\rho_{a}$
of single-crystal  La$_{2}$CuO$_{4+\delta}$ measured at different
values of transport current. In all cases the current was
directed parallel to the crystallographic axis
\textbf{\textit{a}}.  The dependences are presented as $\lg
\rho_{a}$ \textit{vs} $T^{-1/4}$. \vspace{12pt}

Figure 2. The temperature dependence of resistivity $\rho_{a}$ of
single-crystal  La$_{2}$CuO$_{4+\delta}$ measured for transport
current equal to 1~$\mu$A. The current was directed parallel to
the crystallographic axis \textbf{\textit{a}}. The inset shows
temperature behavior of the ratio of the resistivities $\rho_b$
and $\rho_a$ for measuring currents directed along the
crystallographic axes \textbf{\textit{b}} and
\textbf{\textit{a}}. \vspace{12pt}

Figure 3. Magnetoresistance curves at $T=5$~K and $T=20$~K
measured for single-crystal La$_{2}$CuO$_{4+\delta}$ in the
out-of-plane DC magnetic field ($H\|$\textbf{\textit{c}}) for
different amplitudes of measuring current directed along the
crystallographic axis \textbf{\textit{a}}. \vspace{12pt}

Figure 4. Magnetoresistance curves at various fixed temperatures
measured for single-crystal La$_{2}$CuO$_{4+\delta}$ in the
out-of-plane DC magnetic field ($H\|$\textbf{\textit{c}})  for
measuring current (100~$\mu$A) directed along the crystallographic
axis \textbf{\textit{a}}. \vspace{12pt}

Figure 5. Magnetoresistance curves at various fixed temperatures
measured for single-crystal La$_{2}$CuO$_{4+\delta}$ in the
out-of-plane DC magnetic field ($H\|$\textbf{\textit{c}}) for
measuring current (100~$\mu$A) directed along the crystallographic
axis \textbf{\textit{b}}. \vspace{12pt}

Figure 6. Magnetoresistance curves registered for single-crystal
La$_{2}$CuO$_{4+\delta}$  in the out-of-plane pulse magnetic field
($H\|$\textbf{\textit{c}}) at $T=$~4.2 K for the in-plane and
out-of-plane current directions ($J\|$\textbf{\textit{a}} and
$J\|$\textbf{\textit{c}}) with current magnitudes 6 $\mu$A and 7.4
$\mu$A, respectively. \vspace{12pt}

Figure 7. Magnetoresistance curves registered for single-crystal
La$_{2}$CuO$_{4+\delta}$  in the out-of-plane pulse magnetic field
($H\|$\textbf{\textit{c}}) at $T=$~77 K for the in-plane and
out-of-plane current directions ($J\|$\textbf{\textit{a}} and
$J\|$\textbf{\textit{c}}) with current magnitudes 5.93 mA and 178
$\mu$A, respectively. \vspace{12pt}

\newpage
\begin{figure}[htb]
\includegraphics[width=1\linewidth]{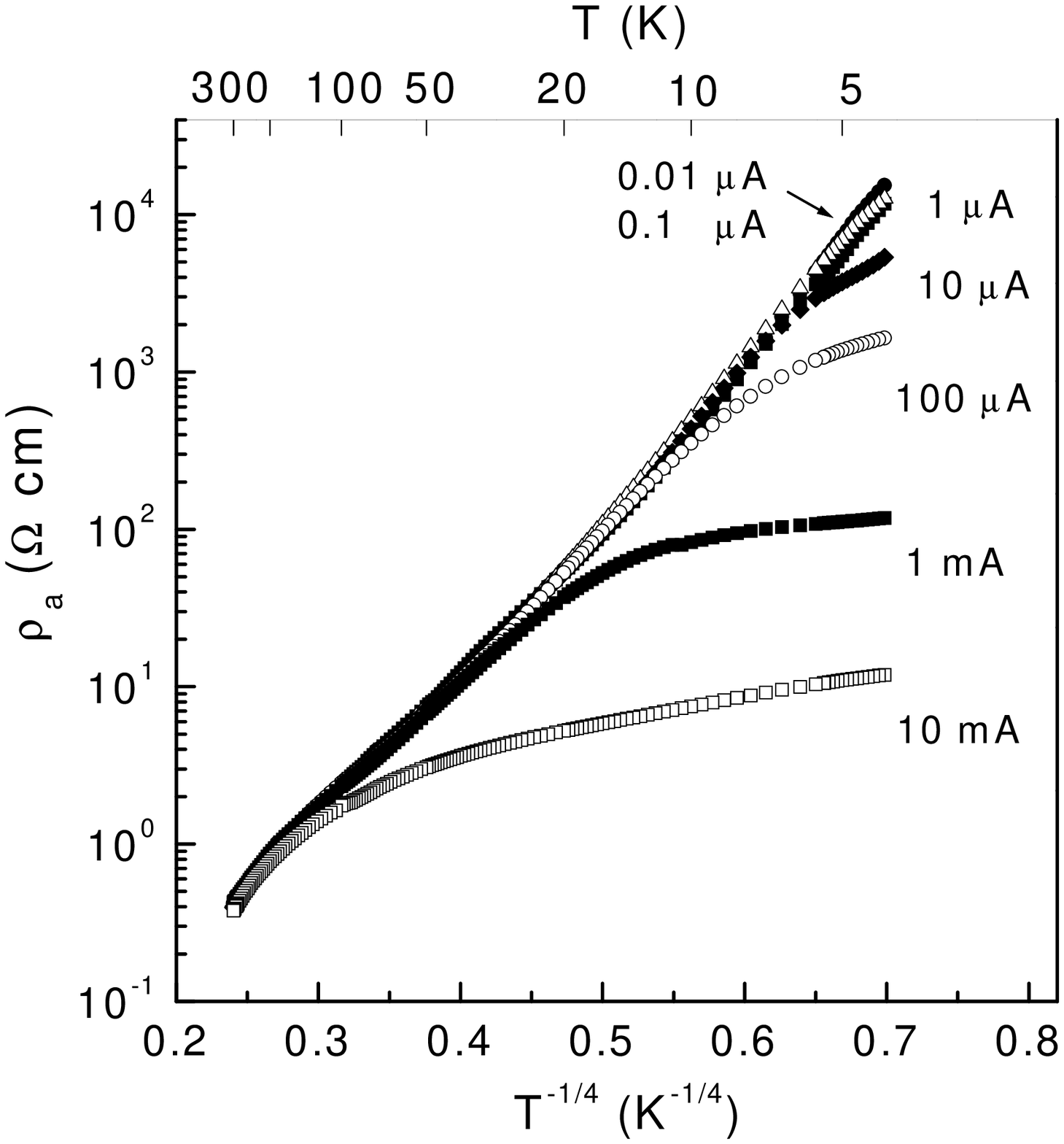}
\end{figure}
\vspace{20pt} \centerline{Figure 1 to paper Belevtsev et al.}

\begin{figure}[htb]
\includegraphics[width=1\linewidth]{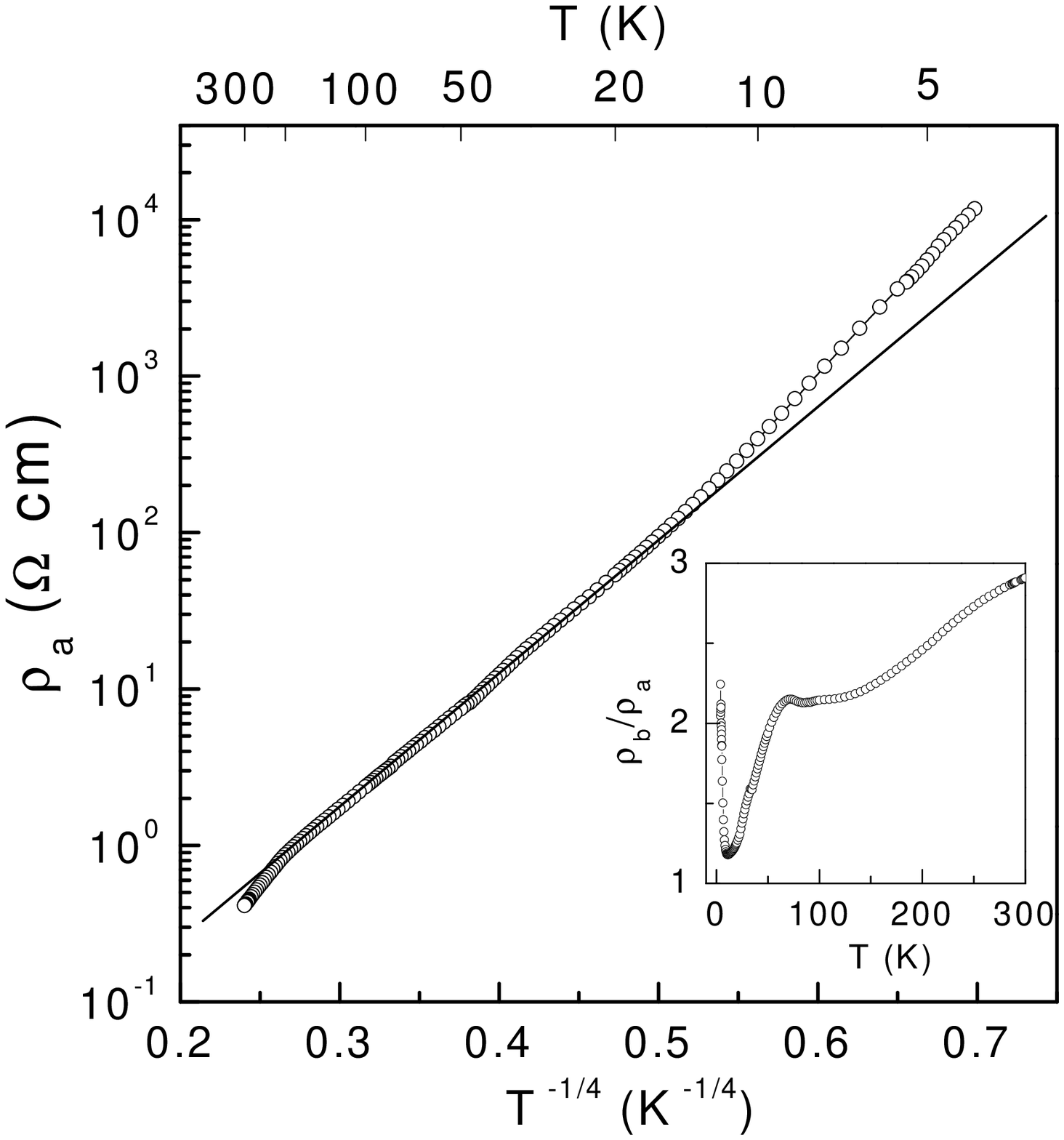}
\end{figure}
\vspace{20pt} \centerline{Figure 2 to paper Belevtsev et al.}

\newpage
\begin{figure}[htb]
\includegraphics[width=1\linewidth]{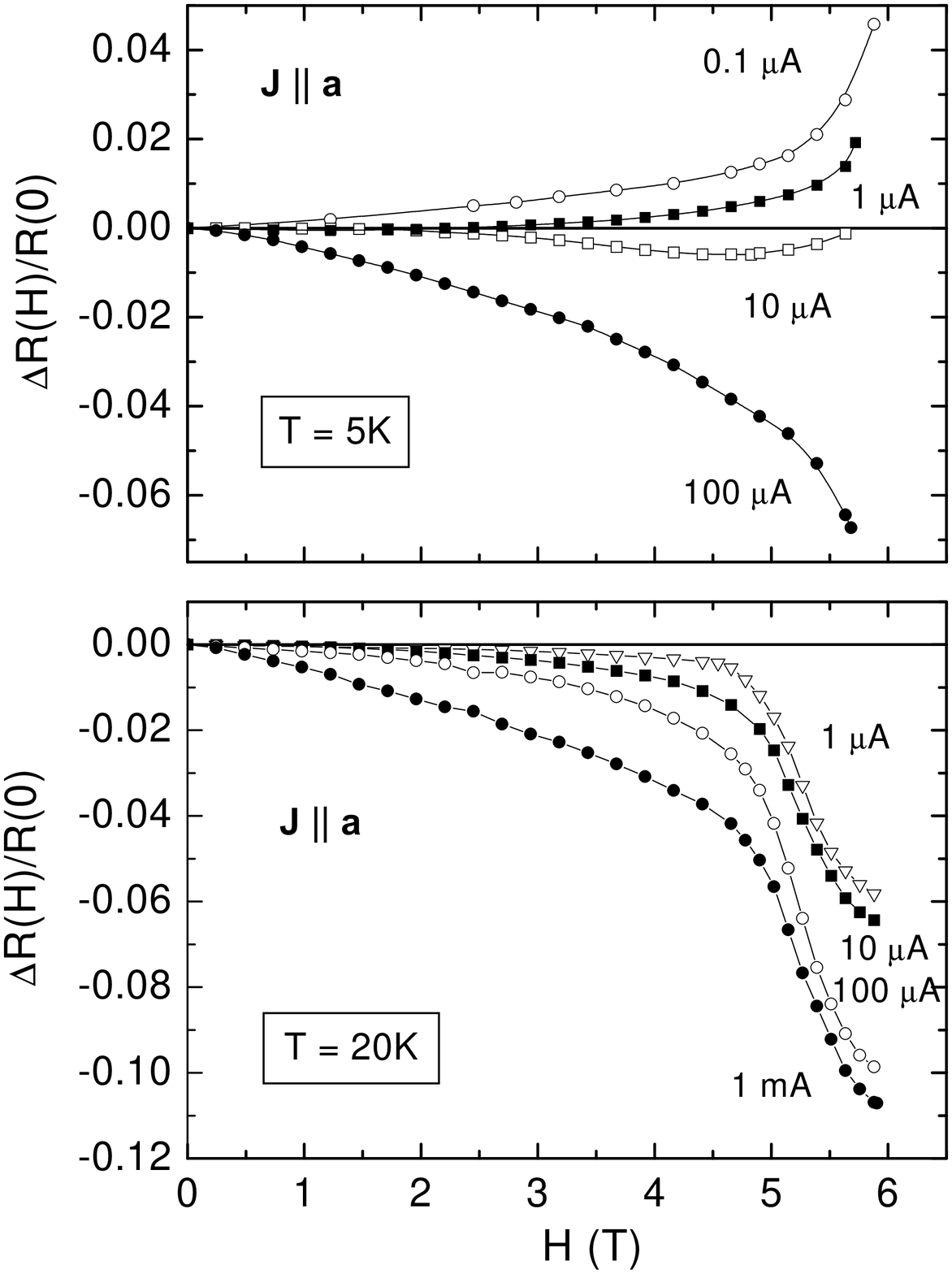}
\end{figure}
\vspace{20pt} \centerline{Figure 3 to paper Belevtsev et al.}

\newpage
\begin{figure}[htb]
\includegraphics[width=1.1\linewidth]{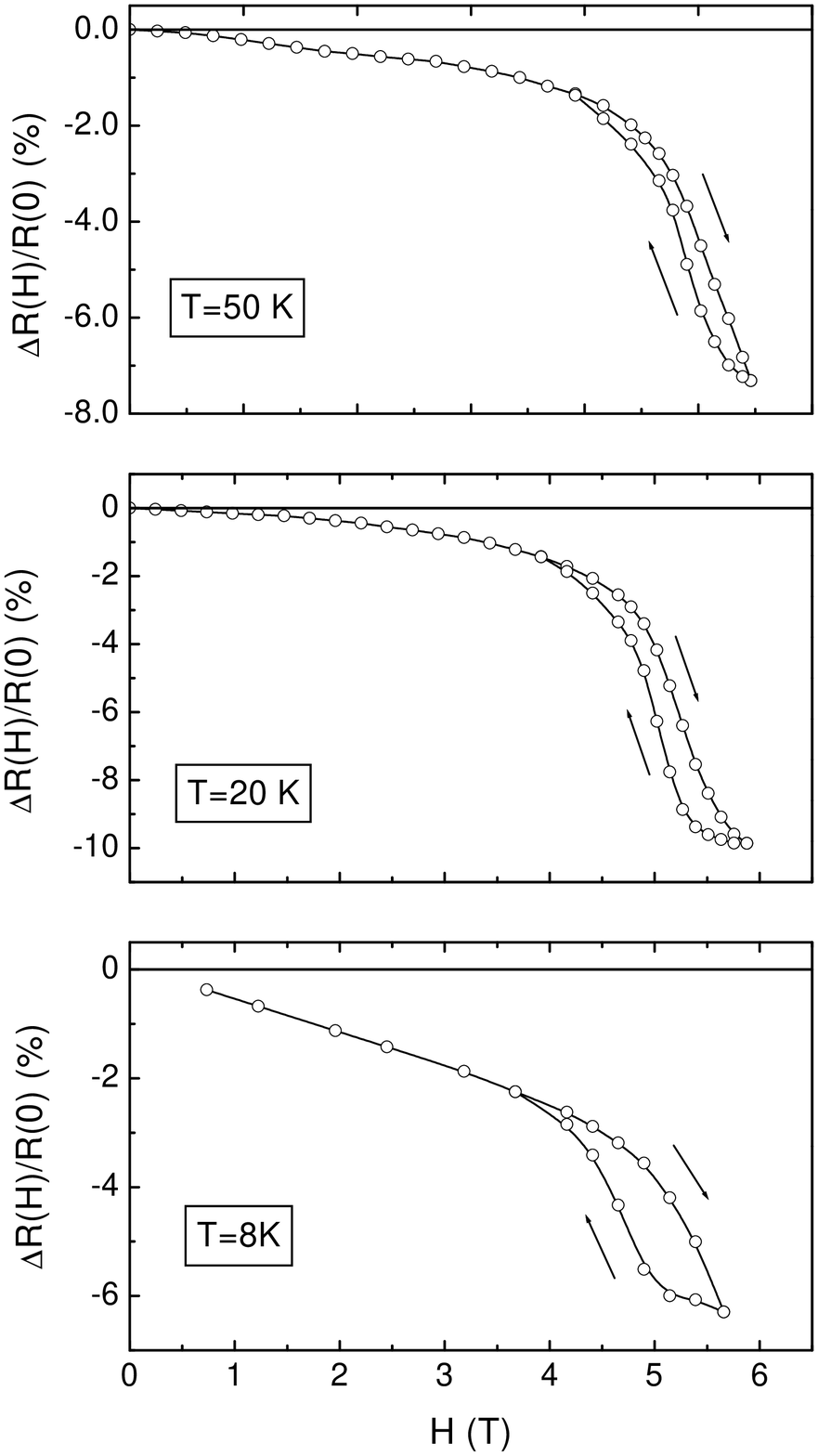}
\end{figure}
\vspace{20pt} \centerline{Figure 4 to paper Belevtsev et al.}

\newpage
\begin{figure}[htb]
\includegraphics[width=1.1\linewidth]{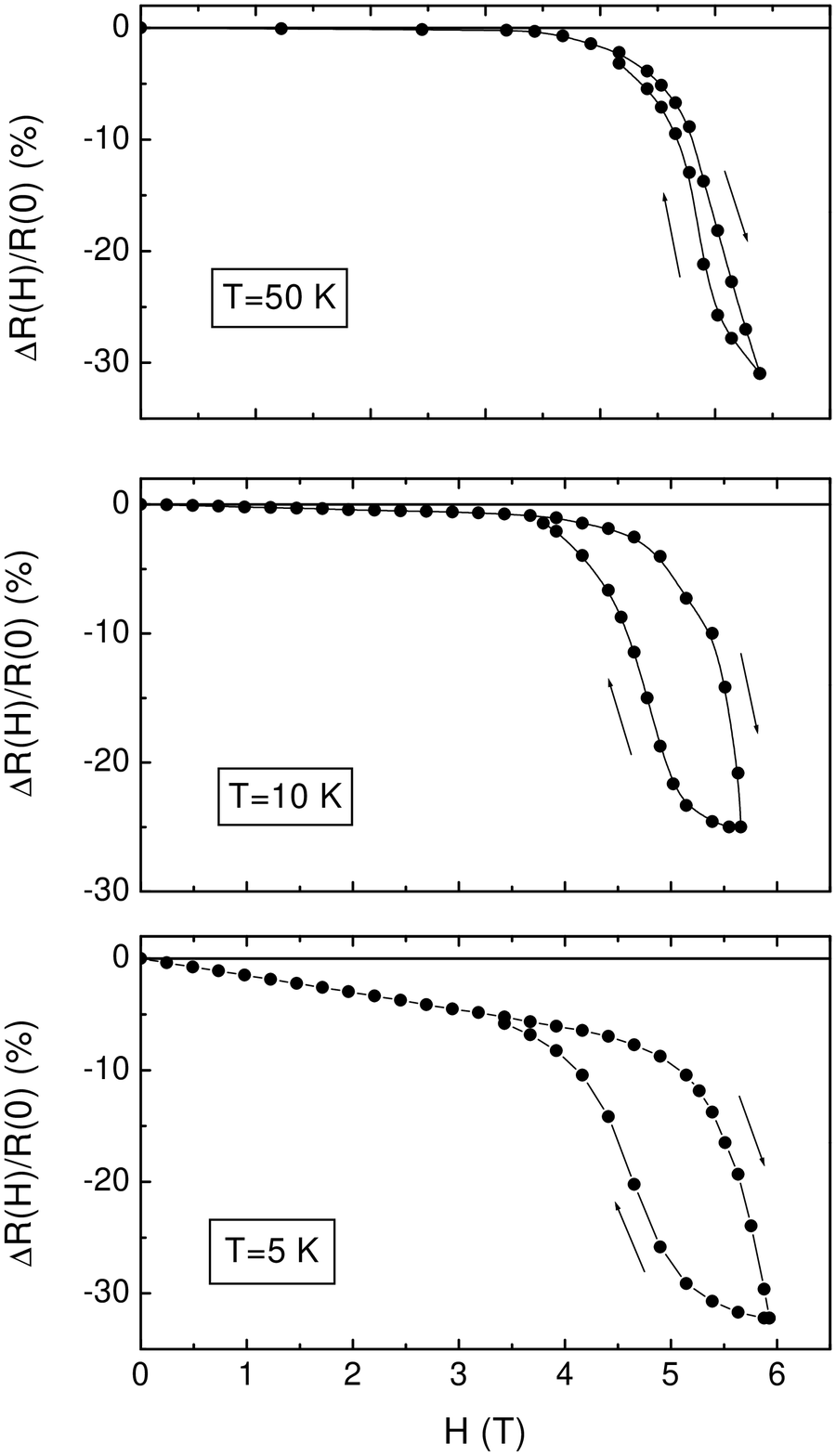}
\end{figure}
\vspace{20pt} \centerline{Figure 5 to paper Belevtsev et al.}

\newpage
\begin{figure}[htb]
\includegraphics[width=1\linewidth]{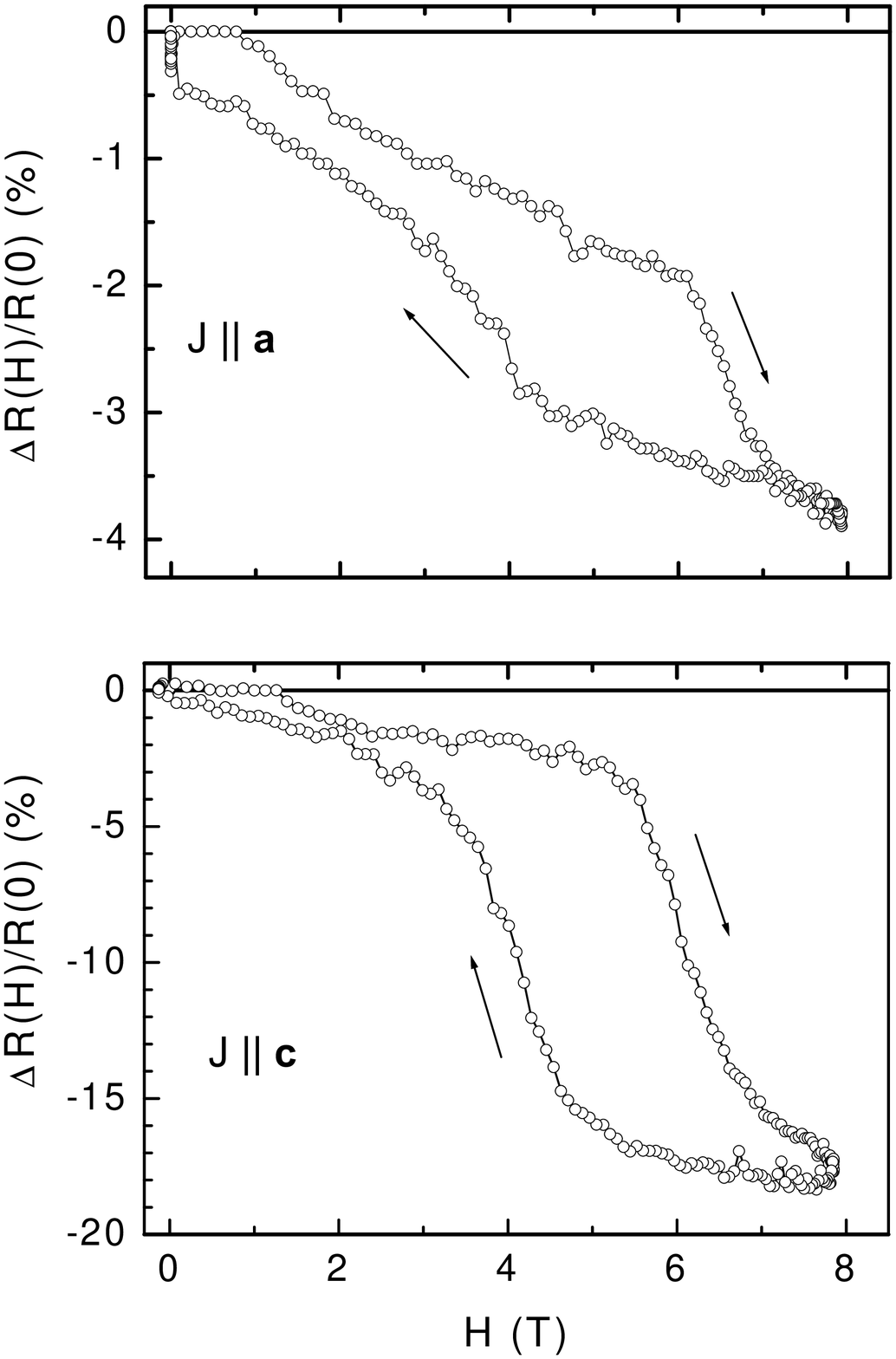}
\end{figure}
\vspace{20pt} \centerline{Figure 6 to paper Belevtsev et al.}

\newpage
\begin{figure}[htb]
\includegraphics[width=1\linewidth]{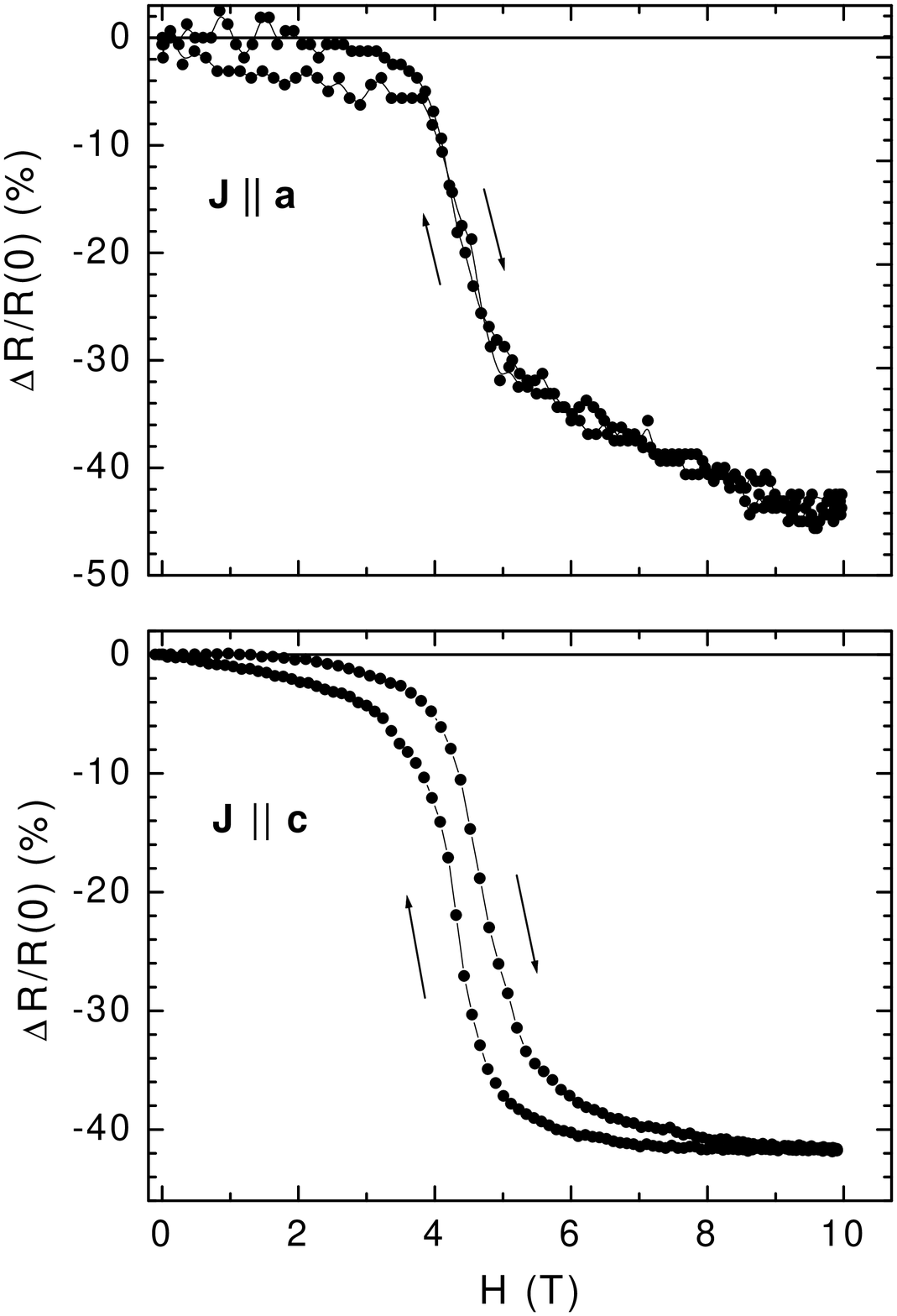}
\end{figure}
\vspace{20pt} \centerline{Figure 7 to paper Belevtsev et al.}
\end{document}